\documentclass[graphics,twocolumn,floatfix,mathbbm,pra,superscriptaddress,aps]{revtex4}
\usepackage{graphicx}
\usepackage{amsmath,amssymb,graphicx,color}
\usepackage{float}
\newcommand{\PrYSO}{Pr$^{3+}$:Y$_2$Si{O$_5$}}
\newcommand{\YSO}{Y$_2$Si{O$_5$}}
\newcommand{\EuYSO}{Eu$^{3+}$:Y$_2$Si{O$_5$}}
 \newcommand{\Prs}{$\text{Pr}^{3+}\text{:Y}_2\text{SiO}_5$ }

\begin{document}

\title{Quantum correlations between single telecom photons and a multimode on-demand solid state quantum memory}

\author{Alessandro Seri}
\author{Andreas Lenhard}
\author{Daniel Riel\"ander}
\author{Mustafa G\"{u}ndo\u{g}an}
\altaffiliation{Present address: Cavendish Laboratory, University of Cambridge, J.J. Thomson Avenue, Cambridge CB3 0HE, United Kingdom}
\author{Patrick M. Ledingham}
\altaffiliation{Present address: Department of Physics, Clarendon Laboratory, University of Oxford, Oxford OX1 3PU, United Kingdom}
\author{Margherita Mazzera}
\email{margherita.mazzera@icfo.es}
\affiliation{ICFO-Institut de Ciencies Fotoniques, The Barcelona Institute of Science and Technology, Mediterranean Technology Park, 08860 Castelldefels (Barcelona), Spain}
\author{Hugues de Riedmatten}
\affiliation{ICFO-Institut de Ciencies Fotoniques, The Barcelona Institute of Science and Technology, Mediterranean Technology Park, 08860 Castelldefels (Barcelona), Spain}
\affiliation{ICREA-Instituci\'{o} Catalana de Recerca i Estudis Avan\c cats, 08015 Barcelona, Spain}

\date{\today}

\begin{abstract} 
Quantum correlations between long lived quantum memories and telecom photons that can propagate with low loss in optical fibers are an essential resource for the realization of large scale quantum information networks. Significant progress has been realized in this direction with atomic and solid state systems. 
Here, we demonstrate quantum correlations between a telecom photon and a multimode on-demand solid state quantum memory. This is achieved by mapping a correlated single photon onto a spin collective excitation in a \PrYSO\, crystal for a controllable time. The stored single photons are generated by cavity enhanced spontaneous parametric down conversion (SPDC) and heralded by their partner photons at telecom wavelength. These results represent the first demonstration of a multimode on-demand solid state quantum memory for external quantum states of light. They provide an important resource for quantum repeaters and pave the way for the implementation of quantum information networks with distant solid-state quantum nodes.
\end{abstract}

\maketitle
\section{INTRODUCTION}

Photonic quantum memories \cite{Afzelius2015} are essential elements for quantum information networks \cite{Kimble2008}, providing efficient and on-demand interfacing between single photons and stationary qubits, e.g. atomic gases \cite{Radnaev2010,Choi2008,Yang2016,Cho2016}, electronic spins in diamonds \cite{Hensen2015,Yang2016a} or phonons \cite{England2015,Riedinger2016}. 
Besides featuring efficiencies and storage times which compare or even overcome those of atomic gases \cite{Hedges2010,Schraft2016,Heinze2013,Zhong2015a}, solid state photonic memories based on rare earth doped crystals are becoming increasingly important as they offer prospects for scalability and integrability \cite{Saglamyurek2011,Zhong2015,Marzban2015,Corrielli2016}. Most protocols for long distance quantum communication require quantum memories connected to communication channels through fibers. One possible direction to achieve this goal is to use telecom quantum memories, e.g. based on erbium doped solids \cite{Lauritzen2010, Saglamyurek2015} or optomechanical systems \cite{Riedinger2016}. However, the most efficient and long lived storage systems up to date are working at wavelengths far from the telecom window, leading to large loss in optical fibers. Possible solutions to overcome this problem include quantum frequency conversion \cite{Radnaev2010,DeGreve2012,Albrecht2014,Maring2014,Farrera2016a,Ikuta2016} or non-degenerate photon pair sources to establish entanglement between quantum memories and telecom photons \cite{Simon2007, Clausen2011, Saglamyurek2011, Rielander2014, Zhang2016}. 
The latter approach has been demonstrated using the atomic frequency comb scheme \cite{Afzelius2009} in rare earth doped single crystals or waveguides \cite{Clausen2011, Saglamyurek2011, Rielander2014}, but the storage of photonic entanglement was performed so far only in the excited state for short and pre-determined storage times. 

Longer and programmable storage times can be obtained by transferring the optical atomic excitations to long lived spin collective excitations (spin waves) thanks to control laser pulses \cite{Afzelius2009}. 
Recently, spin wave storage of weak coherent states at the single photon level \cite{Gundogan2015,Jobez2015}, including qubit storage \cite{Gundogan2015,Laplane2016}, has been demonstrated with rare-earth doped crystals. The generation and storage of continuous variable entanglement between a multimode solid state quantum memory and a light field have also been reported recently \cite{Ferguson2016}. 
In this experiment, light-matter entanglement is created within the memory between spontaneously emitted light and spin waves, the matter part then being converted into a light field. This is a 'read-only' quantum memory \cite{Afzelius2015} with the generated light fields being resonant and, for this demonstration, outside the telecommunication band. This motivates the need for a 'write-read' quantum memory that can store an externally prepared quantum optical state sharing a quantum correlation with a telecom band photon. To date, this has not been achieved with an on-demand spin wave solid state quantum memory. 
Its successful realization with single photons requires an efficient quantum light source matching the spectral properties of the quantum memory \cite{Fekete2013,Rielander2016}, and the quasi-suppression of the noise generated by the strong control pulses. These tasks are challenging due to the small spectral separation between the hyperfine states of the optically active ions (a few MHz in our system). 

\begin{figure*}
\centerline{\includegraphics[width=1.9\columnwidth]{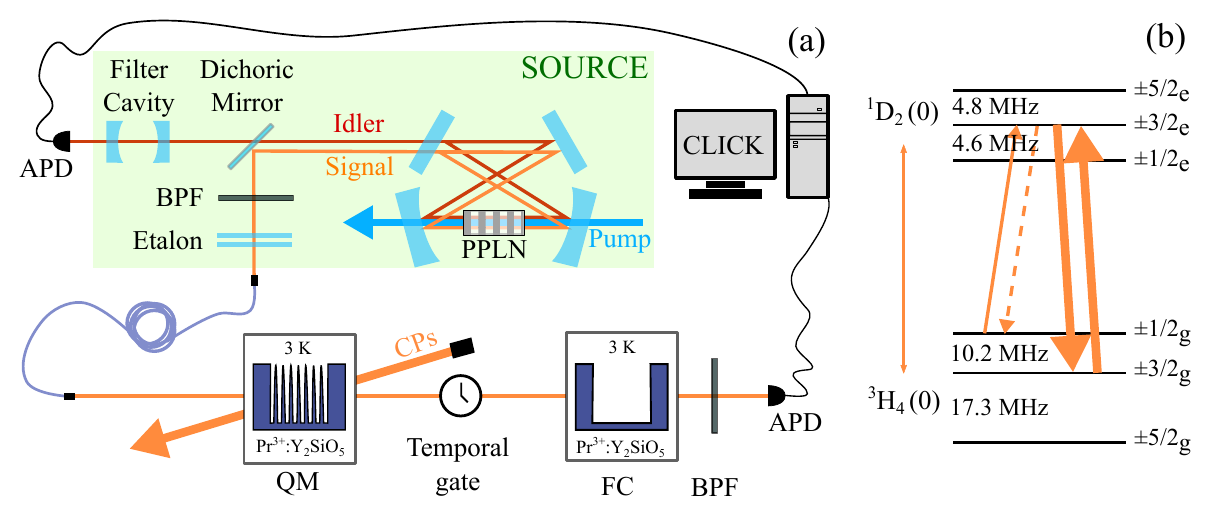}}
\caption{(a) The experimental setup. Photon pairs are created by cavity enhanced SPDC in a  periodically poled lithium niobate (PPLN) crystal. At the output of the cavity they are separated by a  dichroic mirror (DM). The idler photon is further filtered with a cavity (FCav) to select a single frequency mode, before being coupled in an optical fiber and detected with an InGaAs single photon detector (SPD). The signal photon is spectrally filtered with a band-pass filter (BPF) and an etalon, before being coupled in a single mode optical fiber and directed towards the quantum memory (QM), where it is stored using the AFC scheme. Control pulses (CPs) are used to transfer optical excitations to spin-waves and back; they are sent at an angle and counter propagating with respect to the input photons. The retrieved single photon is spectrally filtered using a filter crystal (FC) before being detected by a Silicon SPD. Temporal filtering is achieved with acousto-optic modulators, placed after the memory crystal, opened  only when we expect the SW echo. This prevents additional holeburning in the filter crystal and the SPD to be blinded by eventual leakage of the control pulses. During the preparation of the memory crystal via optical pumping, the SPDs of both idler and signal arms are gated off. After each AFC preparation, the gates are opened and we detect the arrival time of both photons of the pair during a measurement time of about $100\,\mathrm{ms}$, leading to a duty-cycle of 0.14.  (b) Hyperfine splitting of the first sublevels of the ground $^{3}$H$_{4}$ and the excited $^{1}$D$_{2}$ manifold of Pr$^{3+}$ in \YSO. The arrows highlight the $\Lambda$ system chosen for the storage.  
} 
\label{setup}
\end{figure*}

Here, we demonstrate the spin-wave (SW) storage with on-demand retrieval of heralded single photons in a \PrYSO\, crystal using the full atomic frequency comb (AFC) scheme. This is achieved by generating pairs of non-degenerate single photons where one photon is resonant with the optical transition of \PrYSO\, while the other photon is at telecom wavelength. The telecom photon is used to herald the presence of the other photon, which is stored as a SW in the crystal and retrieved on-demand after a controllable time. We measure second order cross correlation values between the heralding and the retrieved photons which exceed the classical bound fixed by the Cauchy-Schwarz inequality for storage times longer than $30\,\mathrm{\mu s}$, effectively demonstrating quantum correlations between telecom photons and single spin waves in a solid. Moreover we demonstrate that our memory can store spin waves in multiple independent temporal modes. 

\section{Experimental details}

\begin{figure*}
\centerline{\includegraphics[width=1.8\columnwidth]{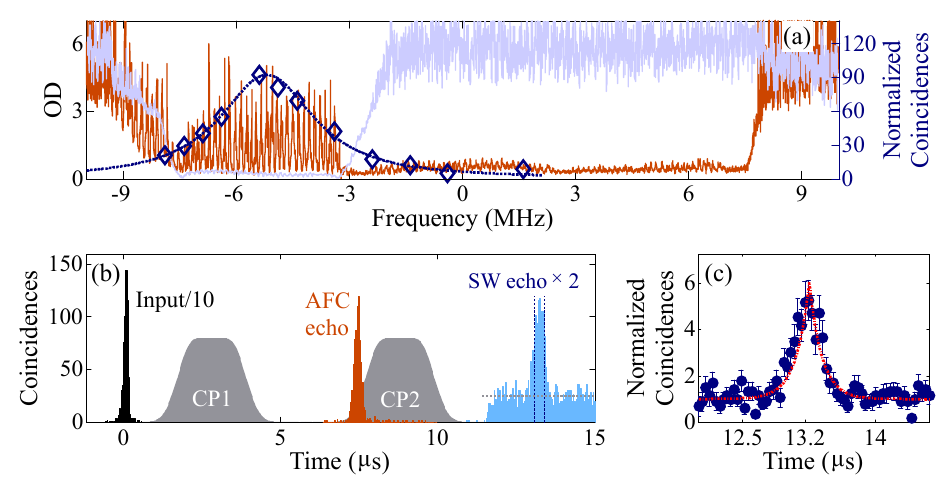}}
\caption{ (a) Example of atomic frequency comb prepared for $\tau = 7.3\,\mathrm{\mu s}$ (red trace). The light violet trace is the transparency window that we burn in the filter crystal. The diamonds are signal-idler coincidence rates taken after preparing a single $800\,\mathrm{kHz}$ broad transparency window in the memory crystal and moving its frequency along the input photons. The error is smaller than the data points. The dotted blue line is a simulation of a Lorentzian peak with $FWHM = 2.8\,\mathrm{MHz}$ convoluted with a $800\,\mathrm{kHz}$-wide spectral hole. 
(b) Time histograms of the input photons (black trace), the AFC echo at $\tau = 7.3\,\mathrm{\mu s}$ (red trace), and the SW echo (blue trace) acquired over an integration time of $343\,\mathrm{min}$. We construct the coincidence histogram by taking the detection of the idler photons as a start and the detection of a signal photon after the memory as stop. The coincidence count rate in the AFC echo is $4 /\mathrm{min}$ and in the SW echo $\approx 1 /\mathrm{min}$. 
The control pulses, detected with a reference photodetector, are displayed as plain pulses and are separated by $T_s = 6\,\mathrm{\mu s}$. They are modulated in amplitude and frequency with Gaussian and hyperbolic tangent waveforms, respectively, as described in the Appendix. The peak power is $21\,\mathrm{mW}$. The dashed vertical lines indicate the integration window for the signal ($\Delta T_d = 320\,\mathrm{ns}$), while the dashed horizontal line represents the noise floor.
(c) Coincidence counts for the SW echo at $T = \tau + T_s = 13.3 \,\mathrm{\mu s}$ normalized by the average noise level, along with its fit to a double exponential function, to account for the Lorentzian spectral shape of the SPDC photons. 
 } 
\label{SNR}
\end{figure*}

Fig. \ref{setup} depicts the experimental setup (a) and the relevant energy level scheme of \PrYSO\, (b) where the chosen $\Lambda$ system is indicated by arrows. The atomic frequency comb is prepared, following the spectral hole burning procedure described in \cite{Jobez2016}, at the frequency of the ${1}/{2}_\textrm{g}-{3}/{2}_\textrm{e}$ transition and the control pulses drive the coherence from the ${3}/{2}_\textrm{e}$ to the empty ${3}/{2}_\textrm{g}$ storage state.
The narrow-band spectral filtering of the noise resulting from the control pulses is accomplished with a second \PrYSO\, crystal, where a narrow transparency window (around $5.5\,\mathrm{MHz}$) is burned at the frequency of the AFC \cite{Gundogan2015}. An example of AFC prepared in the memory crystal overlapped with the narrow spectral hole burned in the filter crystal is reported in Fig. \ref{SNR}(a). 

Our cavity-enhanced SPDC source produces ultra-narrow photon pairs where one photon, the idler, is in the telecom E-band at $1436\,\mathrm{nm}$, and the other is resonant with the Pr$^{3+}$ optical transition at $606 \,\mathrm{nm}$, specifically with the transition where the AFC is prepared \cite{Fekete2013,Rielander2014}. The pump frequency is $426.2\,\mathrm{nm}$ and the average power for the measurements presented in this paper is 3.3 $\pm$ 0.5 mW.
The probability to obtain a single signal photon in front of the quantum memory conditioned on a detection in the idler SPD (i.e. the heralding efficiency) is $\eta_H= (20.9 \pm 0.5) \%$. We also switch off the SPDC pump after the detection of the idler photons for $30$ to $40\,\mathrm{\mu s}$ (depending on the experiment), thus interrupting the creation of photon pairs during the detection of the stored and retrieved photons. 
Further details about the experimental setup and the preparation of the atomic frequency comb can be found in the Appendix. 

\section{Heralded single photon spectrum}
\label{SP}
To access the spectrum of the heralded photons to store we compare two measurements. First we measure the temporal distribution of coincidences between the signal and idler photons of the source alone \cite{Rielander2016}. This correlation function has a width of $78\,\mathrm{ns}$ ($FWHM$) which is the correlation time of the photons. From this value we calculate a biphoton bandwidth of $2.8\,\mathrm{MHz}$.
In a second experiment we employ the \PrYSO\, memory crystal as a tunable frequency filter \cite{Zhang2012,Beavan2013,Gundogan2015}. We prepare a $800\,\mathrm{kHz}$-wide spectral hole and record coincidence histograms when the central frequency is swept by about $10\,\mathrm{MHz}$ around the frequency of the signal photons. In this experiment the photons passing through the transparency window are directly steered to the APD for detection, bypassing the temporal and spectral filtering stages. The coincidence rate as a function of the hole position (blue diamonds overlapped to the AFC in Fig. \ref{SNR}(a)) gives a measure of the spectral distribution of the heralded single photons at $606\,\mathrm{nm}$. The result of this measurement agrees with the spectrum extrapolated from the signal-idler coincidence histogram
measured immediately after the SPDC source \cite{Rielander2016}. This is confirmed by the good overlap between the diamonds and the blue dotted line, which represents the convolution of a Lorentzian curve of width $2.8\,\mathrm{MHz}$ and the trace of the $800\,\mathrm{kHz}$-wide spectral hole. We estimate the spectral overlap between the heralded single photons and the AFC to be about $70\,\%$, which currently limits the AFC storage efficiency. 

\section{Spin wave storage of heralded single photons}
\label{SWS}
We first measure the coincidence histogram when the signal photons are sent through a $18\,\mathrm{MHz}$-wide transparency window prepared in the memory crystal (input trace at $0\,\mathrm{\mu s}$ in Fig. \ref{SNR}(b)). The correlation time between the signal and idler photons is $\tau_c =(89\pm 4)\,\mathrm{ns}$, leading to a heralded photon linewidth of $(2.5 \pm 0.1)\,\mathrm{MHz}$ \cite{Rielander2016}. The correlation between the two photons is inferred by measuring the normalized second order cross-correlation function $g^{(2)}_{s,i} = {P_{s,i}}/({P_{s} \cdot P_i})$, where $P_{s,i}$ is the probability to detect a coincidence between idler and signal photons in a time window $\Delta T_d = 320\,\mathrm{ns}$, while $P_{i}$ and $P_{s}$ are the uncorrelated probabilities to detect each photon. We find a $g^{(2)}_{s,i}$ value of $96 \pm 32$. Then we prepare an AFC with periodicity $\Delta$ and store the single photon as a collective optical atomic excitation $\left| \psi_e \right\rangle = \frac{1}{\sqrt{N}} \sum_{j=1}^{N} e^{i\textbf{x}_j\cdot \textbf{k}_{in} } \left|g_1 \ldots e_j \ldots g_N\right\rangle,$ where $ \textbf{k}_{in}$ is the wave-vector of the incoming photon (see level scheme in Fig. \ref{setup}(b)). We obtain, for a pre-programmed storage time $\tau = \frac{1}{\Delta} = 7.3\,\mathrm{\mu s}$, an efficiency $\eta_{AFC} = (11.0 \pm 0.5)\,\%$ and $g^{(2)}_{AFC,i}= 130 \pm 31$  (see Fig. \ref{SNR}(b)). This result represents an improvement in terms of $g^{(2)}_{AFC,i}$ of more than one order of magnitude compared to the state of the art in the same system \cite{Rielander2014}. The $g^{(2)}$ value increases after the AFC storage due to the fact that the stored photons are transferred to a temporal mode free of noise \cite{Rielander2014}. After the retrieval, we measure $\tau_c= (147 \pm 7) \,\mathrm{ns}$, larger than the value measured before storage. We attribute this temporal stretching to spectral mismatch between the input photons before the memory ($FWHM = 2.8\,\mathrm{MHz}$) and the atomic frequency comb (total width $4\,\mathrm{MHz}$), as evidenced in Fig. \ref{SNR}(a). 

We then perform spin wave storage experiments by sending pairs of strong control pulses after the detection of each heralding photon. The first control pulse with wave vector $\textbf{k}_{C}$ transfers the collective optical excitation $\left| \psi_e \right\rangle$ to a collective spin excitation (a spin wave), which can be written as $\left| \psi_{sw} \right\rangle = \frac{1}{\sqrt{N}} \sum_{j=1}^{N} e^{i\textbf{x}_j\cdot (\textbf{k}_{in}- \textbf{k}_{C}}) \left|g_1 \ldots s_j \ldots g_N\right\rangle.$
For a storage time in the ground state of $T_s = 6\,\mathrm{\mu s}$ (thus a total storage time $T = T_s + \tau = 13.3\,\mathrm{\mu s}$), we detect the retrieved photons, i.e. the spin-wave echo ($swe$), with an efficiency $\eta_{sw} = (3.6 \pm 0.2) \,\%$. $\eta_{sw}$ is inferred with a coincidence window of  $\Delta T_d = 320\,\mathrm{ns}$, containing $80\,\%$ of the signal and including the loss in the filter crystal. Fig. \ref{SNR}(c) shows a magnification of the SW echo mode, normalized by the average value of the noise measured outside the peak, leading directly to the signal-to-noise ratio (SNR) of the stored and retrieved photons. We observe a maximum SNR of around 5. This curve can also be used to infer $g^{(2)}_{swe,i}$. With our filtering strategy we reach a noise floor of $(2.0 \pm 0.1) \times 10^{-3}$ photons per storage trial at the memory crystal (horizontal dashed line in the SW echo temporal mode). 
The correlation time of $\tau_c= (200 \pm 40) \,\mathrm{ns}$ exceeds the one after storage in the excited state. This further increase after the SW storage is attributed to the limited chirp of the control pulses (see Appendix).  

\section{Quantum correlation between single telecom photons and single spin waves}

\begin{figure*}
\centering\includegraphics[width=1.8\columnwidth]{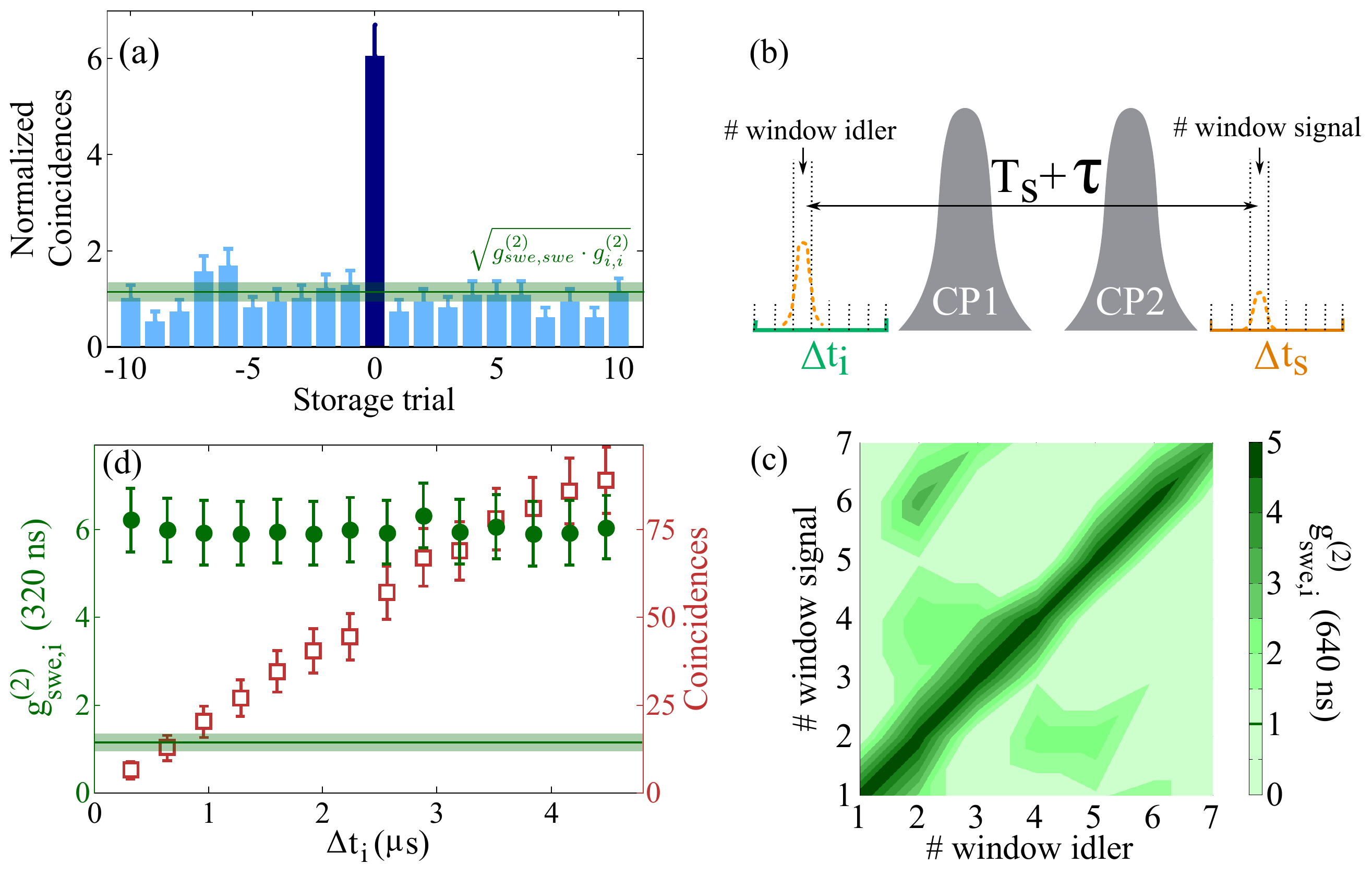}
\caption{ (a) Unconditional cross-correlation between the idler photons and the retrieved signal photons. The $g^{(2)}_{swe,i}(\Delta T_d = 320\, \mathrm{ns})$ value for this measurement is $6.1 \pm 0.7$. The classical bound given by the Cauchy-Schwarz inequality is reported as a horizontal line. The error bars are calculated considering Poissonian statistics. (b) Scheme representing the analysis of the cross-correlation measurement to evidence the multimodality. Both idler and retrieved signal integration windows are divided into smaller intervals ($640\,\mathrm{ns}$) and the correlation is calculated between intervals separated by different storage time. (c) Cross-correlation value between idler and retrieved signal photons detected in small temporal modes separated by different storage times. For this analysis, we also set $\Delta T_d = 640\,\mathrm{ns}$ for better statistics. The $g^{(2)}_{swe,i}(\Delta T_d = 640\, \mathrm{ns})$ exceeds the classical threshold (represented in the color code bar for $\Delta T_d = 640\, \mathrm{ns}$) only for windows separated by the total storage time $T = 13.3\,\mathrm{\mu s}$. Violations outside these windows are not statistically significant due to a low number of counts. (d) Cross-correlation value between the idler photons and the retrieved signal photons (full circles) and the coincidence counts in the SW echo (empty squares) as a function of the detection window for the idler photons. Both the signal detection window and the coincidence window remain constant at $4.5\,\mathrm{\mu s}$ and $320 \,\mathrm{ns}$, respectively. The classical threshold is also reported as a horizontal line. The integration time for this measurement is 38.5 h. The error bars are calculated considering Poissonian statistics. 
}
\label{multimodality}
\end{figure*}

To investigate the non-classical nature of the photon correlation after the SW storage, we measure  $g^{(2)}_{swe,i} (\Delta T_d)$ and compare it to the unconditional autocorrelation of the idler ($g^{(2)}_{i,i}(\Delta T_d)$) and retrieved signal ($g^{(2)}_{swe,swe}(\Delta T_d)$) fields, respectively. To access these quantities we correlate photon detections from different unconditional storage iterations (see Fig. \ref{multimodality}(a)). The time between two consecutive storage trials is $190\,\mathrm{\mu s}$. In each trial (500 per comb preparation), we maintain the gate of the idler SPD open during $4.5\,\mathrm{\mu s}$ before sending the control pulses. We find $g^{(2)}_{swe,i} = 6.1 \pm 0.7 $. We note that our $g^{(2)}_{swe,i}$ is limited by the spin wave read-out efficiency, that we estimate to be approximately $\eta_R = 24\,\%$, while the write efficiency is $\eta_W = 31\,\%$ (see Appendix for a detailed discussion). 
With this unconditional sequence, the measured noise floor ($(1.3 \pm 0.1) \times 10^{-3}$ photons per storage trial) is lower than with the conditional one (section \ref{SWS}). We attribute this result to the fact that the number of control pulse pairs per comb is larger in the unconditional sequence. This probably contributes to a further emptying of the spin storage state. The classical bound from the Cauchy-Schwarz inequality $g^{(2)}_{swe,i} < \sqrt{g^{(2)}_{swe,swe} \cdot g^{(2)}_{i,i}}\,$ is indicated in Fig. \ref{multimodality}(a) as a horizontal line. The measured unconditional autocorrelations are (also for $\Delta T_d = 320\,\mathrm{ns}$) $g^{(2)}_{i,i} = 1.32 \pm 0.04$ \cite{Rielander2016} and $g^{(2)}_{swe,swe} = 1.0 \pm 0.4$ (see Appendix).
The Cauchy-Schwarz parameter $R = \frac{({g^{(2)}_{swe,i}})^2}{g^{(2)}_{swe,swe} \cdot g^{(2)}_{i,i}} = 28 \pm 12$, exceeds the classical limit of $R = 1$ by more than 2 standard deviations. 
The confidence level for violating the Cauchy-Schwarz inequality, i.e. for observing a non-classical correlation between the telecom heralding photon and the single spin wave stored in the crystal, is $98.8\,\%$. If a larger coincidence window is considered, $\Delta T_d = 1\,\mathrm{\mu s}$, the $R$ value is reduced to $8.3 \pm 2.3$ due to a bigger contribution of noise in the $g^{(2)}_{swe,i}$. On the other hand, due to a reduction of the statistical error, the confidence level for the demonstration of non-classical correlation rises up to $99.92\,\%$ (see Appendix).

The main advantage of the full AFC protocol is the possibility to store multiple distinguishable temporal modes, while maintaining their coherence and quantum correlation \cite{Afzelius2009}. This ability is crucial for applications in quantum information protocols, e.g. to enable temporally multiplexed quantum repeater protocols with high communication speed \cite{Simon2007} and storage of time-bin qubits robust against decoherence in optical fibers. To test this aspect, we perform experiments with detection gates much longer than the photons duration. 
We divide both the idler and the retrieved signal detection windows, $\Delta t_i$ and $\Delta t_s$, respectively, into smaller temporal modes of width $640\,\mathrm{ns}$, as sketched in Fig. \ref{multimodality}(b). For this analysis we also consider $\Delta T_d = 640\,\mathrm{ns}$ in order to have better statistics. We then check that we have non-classical correlations between modes separated by the total storage time $T = T_s + \tau = 13.3\,\mathrm{\mu s}$ and classical correlations between modes separated by $T \ne 13.3\,\mathrm{\mu s}$, as shown in Fig. \ref{multimodality}(c). Contrary to other temporally multimode storage protocols \cite{Hedges2010, Sekatski2011, Ferguson2016}, in the full AFC protocol the total storage time is maintained for the different temporal modes. 
To additionally demonstrate that the multimode capacity does not imply any increase of the noise, we compute $g^{(2)}_{swe,i}(\Delta T_d = 320\, \mathrm{ns})$ for idler detection windows $\Delta t_i$ varying from $320\,\mathrm{ns}$ to $ 4.5\,\mathrm{\mu s}$, as shown in Fig. \ref{multimodality}(d) (full circles) together with the coincidence counts measured in the center peak for each window size (empty squares). As expected, the latter increases by increasing $\Delta t_i$ but the $g^{(2)}_{swe,i}$ value remains constant, within the error bar, and well above the classical bound over the whole range. Defining the number of temporal modes $N_m$ as $N_m=\Delta t_i/\Delta T_d$, we confirm non-classical storage of a maximum of 7 temporal modes
with a $\Delta T_d = 640\,\mathrm{ns}$ that contains the $94\,\%$ of the coincidence peak. However, considering $\Delta T_d = 320\,\mathrm{ns}$, which still contains the $80\,\%$ of the SW echo, a $4.5\,\mathrm{\mu s}$-wide gate can accommodate up to 14 independent temporal modes.

\begin{figure}[h]
\centering\includegraphics[width=1\columnwidth]{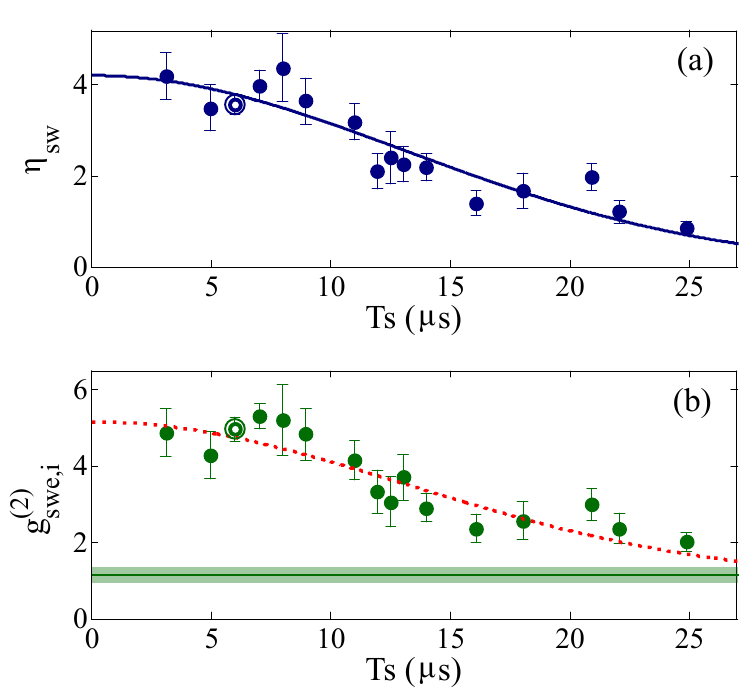}
\caption{(a) Spin-wave storage efficiency as a function of the storage time $T_s$. The solid line is a fit of the experimental data taking into account a Gaussian inhomogeneous broadening of the spin state. The spin inhomogeneous broadening extracted from the fit is $\gamma_{inh} = (20 \pm 3) \,\mathrm{kHz}$. (b) $g^{(2)}_{swe,i}(320\,\mathrm{ns}) $ value as a function of the storage time. The classical threshold is reported as horizontal line. 
In both panels, the error bars are calculated considering Poissonian statistics and the circled data points refer to the measurement reported in Fig. \ref{SNR}. }
\label{storagetime}
\end{figure}

Finally, to illustrate the ability to read-out the stored spin wave on-demand, we perform storage experiments at different SW storage times. For these measurements, we implement a semi-conditional storage sequence in order to obtain good statistics with shorter integration time. We wait for heralding photons and, after each detection, we send 15 pairs of control pulses (each pair being separated by $190 \,\mathrm{\mu s}$ from the neighboring), that we exploit to correlate the idler and the retrieved signal  (see Appendix).
The measured storage and retrieval efficiencies are reported in Fig. \ref{storagetime}(a) together with a Gaussian fit, which accounts for the inhomogeneous broadening of the spin-state. As a fitting parameter we obtain the spin inhomogeneous line width $\gamma_{inh} = (20 \pm 3)\, \mathrm{kHz}$, in good agreement with that measured in different experiments on the same crystal \cite{Gundogan2015,Corrielli2016}. This further confirms that the photons are stored as spin waves. The second order cross-correlation function for increasing SW storage times $T_s$ is shown in panel (b) of Fig. \ref{storagetime}. The red dashed curve is calculated by considering the Gaussian fit of the signal decay (solid curve in panel (a)), normalized by the source heralding efficiency $\eta_H$ and the average noise floor of $(1.9 \pm 0.2) \times 10^{-3}$ photons per trial (the average being over the different $T_s$). We measure non-classical correlations between the idler and the retrieved signal photons up to a total storage time $T = \tau + T_s = 32.3\,\mathrm{\mu s}$. Note that, while we measure the Cauchy-Schwarz parameter $R$ with unconditional measurements at $T = 13.3\,\mathrm{\mu s}$, to assess the non-classicality for longer storage times we make the assumption that the retrieved signal autocorrelation does not increase for semi-conditional measurements at different storage times. This is a conservative estimate, since the $g^{(2)}_{swe,swe}$ value is mainly determined by the noise in the read-out, that we verify being constant over the whole range of storage times investigated and not bunched (see Appendix).
Under this hypothesis, the Cauchy-Schwarz inequality is violated at $T = 32.3\,\mathrm{\mu s}$ with a confidence of $94\,\%$ for $\Delta T_d = 320\,\mathrm{ns}$. 

\section{Discussion and conclusion}
The demonstrated quantum correlation between a telecom photon and a spin wave in a solid is an essential resource to generate entanglement between remote solid state quantum memories \cite{Simon2007}. The measured value of $g^{(2)}_{swe,i}$ after spin wave storage is currently limited by the signal-to-noise ratio of the retrieved photon, which is in turn mostly limited by the low storage and retrieval efficiency. This could be greatly improved by using higher optical depth \cite{Hedges2010} with higher quality combs, or crystals in impedance matched cavities \cite{Sabooni2013,Jobez2014}. The storage time is currently limited by the spin inhomogeneous broadening and could be increased using spin-echo and dynamical decoupling techniques \cite{Jobez2015}, with prospect for achieving values up to one minute \cite{Heinze2013} in our crystal, while even longer storage times (of order of hours) may be available in \EuYSO \cite{Zhong2015a}. Finally our experiment could be extended to the storage of entangled qubits, e.g. using time-bin encoding \cite{Gundogan2015}.  
  
In conclusion, we have reported the first demonstration of quantum storage of heralded single photons in an on-demand solid state quantum memory. We have shown that the non-classical correlations between the heralding and the stored photons are maintained after the retrieval, thus demonstrating non-classical correlations between single telecom photons and single collective spin excitations in a solid. Finally we showed that the full atomic frequency comb protocol employed allows one to store a single photon in multiple independent temporal modes. These results represent a fundamental step towards the implementation of quantum communication networks where solid state quantum memories are interfaced with the current fiber networks operating in the telecom window \cite{Simon2007}. 

\textbf{Acknowledgments.} We acknowledge financial support by the ERC Starting Grant QuLIMA, by the Spanish Ministry of Economy and Competitiveness (MINECO) and Fondo Europeo de Desarrollo Regional (FEDER) (FIS2015-69535-R), by MINECO Severo Ochoa through Grant No. SEV-2015-0522 and through the Ph.D. Fellowship Program (for A.S.), by AGAUR via 2014 SGR 1554, by Fundaci\'o Cellex, and by CERCA Programme/Generalitat de Catalunya.

\newpage

\section{APPENDIX}
\label{app}

In the present Appendix we provide details about the experimental setup (section \ref{Setup}), the preparation and characterization of the atomic frequency comb, including an analysis of the efficiency (in section \ref{fullAFC}), the sequences employed to reconstruct the second order cross- and auto-correlation functions (see section \ref{sequences}).

\section{Setup}
\label{Setup}

\begin{figure}
   \centering
   \includegraphics[width=0.5\textwidth]{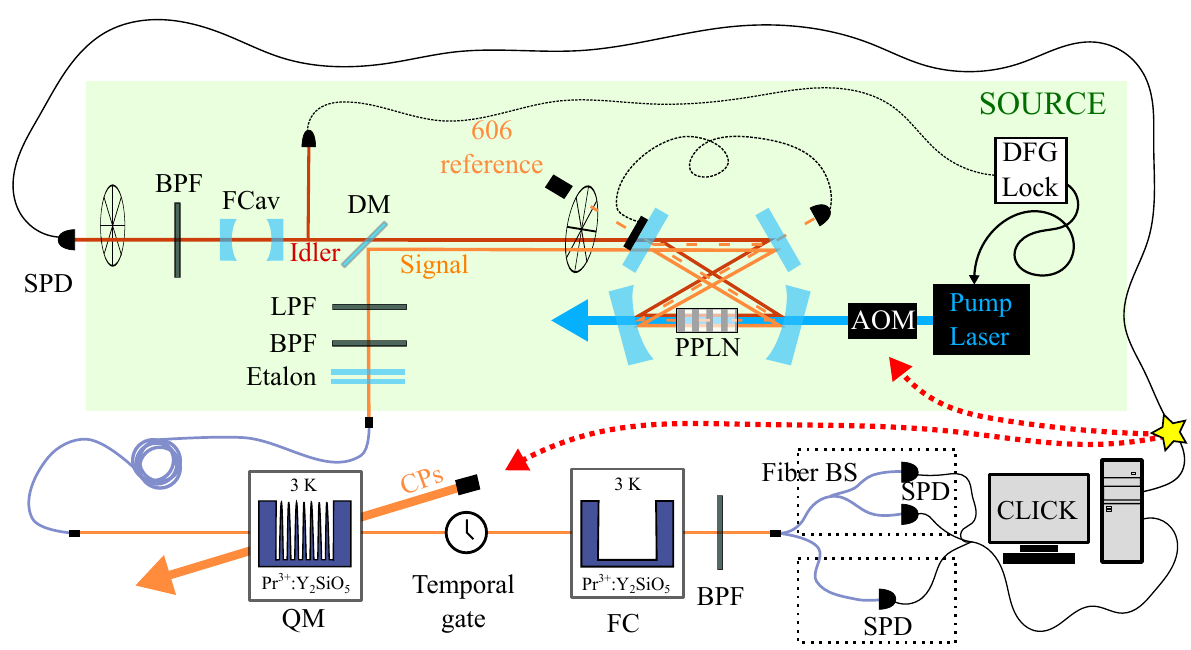}
   \caption{Detailed optical setup of the experiment. The single photon pairs are generated in a bow-tie cavity via SPDC. While the idler photon is measured and used as herald, the signal photon is sent through a fiber to the setup where the storage protocol is performed. DFG stands for difference frequency generation, PPLN for periodically poled lithium niobate, DM for dichroic mirror, FCav for filter cavity, BPF for band pass filter, LPF for low pass filter, QM and FC are the crystals used as quantum memory and filter crystal, respectively, which are placed inside a cryostat. SPD labels the single photon detectors.}
\label{fig:setup}
\end{figure}

The experimental setup, shown in Fig. \ref{fig:setup}, is composed of two main parts, the quantum light source and the quantum memory. 

\subsubsection{Source}

The quantum light source is based on cavity-enhanced spontaneous parametric down conversion (SPDC) \cite{Fekete2013}. A pump at $426.2\,\mathrm{nm}$, whose power is maintained in the range $3-5\,\mathrm{mW}$, produces widely non-degenerate photon pairs when passing through a periodically poled lithium niobate (PPLN) crystal. One photon of the pair, the idler, is at telecom wavelength (Telecom E-band), namely $1436\,\mathrm{nm}$, while the other, the signal, is at $606\,\mathrm{nm}$. To ensure that the signal photons are resonant with the quantum memory, the length of the SPDC cavity is locked to a  reference beam. 
To guarantee simultaneous resonance of the idler photons we measure the light generated by DFG (difference frequency generation) of the pump and the $606\,\mathrm{nm}$ reference beam to derive an error signal which we feed back to the pump frequency.
A mechanical chopper placed before the SPDC cavity enables to alternate the locking and measurement periods, in order to not blind the detection of the single photons with the classical reference beams, the duty cycle being about $45\%$.
The double resonance of widely non-degenerate signal and idler allows for an efficient suppression of the redundant modes of the cavity due to a clustering effect \cite{Pomarico2012,Fekete2013}. At the cavity output the measured spectrum consists of four effective modes in the main cluster (separated by the cavity free spectral range, $423\,\mathrm{MHz}$) \cite{Rielander2016}. To operate in single-mode regime, the idler photons are sent through a homemade filtering cavity (linewidth $80\,\mathrm{MHz}$, free spectral range $16.8\,\mathrm{GHz}$), then coupled into a single mode fiber to the single photon detector (ID230, IDQuantique). This ensures a single-mode herald. The signal photons are filtered with an etalon, which suppresses the side clusters ($44.5\,\mathrm{GHz}$ away from the main cluster), and a bandpass filter (centered at $600\,\mathrm{nm}$, linewidth $10 \,\mathrm{nm}$, Semrock) before being sent to the quantum memory. The second-order cross-correlation function between signal and idler photons measured  before the memory for a coincidence window of $\Delta T_d = 320 \,\mathrm{ns}$ is $g^{(2)}_{s,i} = 57 \pm 1$ (pump power $3.4\,\mathrm{mW}$).

\subsubsection{Quantum Memory setup}

The laser that we use for $606\,\mathrm{nm}$ light is a Toptica DL SHG pro, stabilized with the Pound-Drever-Hall technique to a home-made Fabry-Perot cavity in vacuum \cite{Rielander2014}. From this laser we derive the reference beam for the source and all beams necessary to prepare and operate the memory. The single photons at $606\,\mathrm{nm}$ generated by the source pass through our quantum memory, a \Prs crystal cooled down to $\sim2.7\,\mathrm{K}$ in a cryostat (closed-cycle cryocooler, Oxford Instruments), with a waist of $45\,\mathrm{\mu m}$. The beam for memory preparation and control pulses has a waist of $175\,\mathrm{\mu m}$, and is sent with a small angle with respect to the photons and counter propagating. In this way we can spatially filter out some noise generated by the control pulses.
To protect the single photon counter (SPD) we add a temporal gate which is composed of two AOMs, one aligned in the +1 order, and the other in the -1, to maintain the frequency fixed. The spectral narrow band filter is performed using a second \Prs crystal where we optically pump a $5.5\,\mathrm{MHz}$ spectral hole centered at the frequency of the single photons \cite{Gundogan2015}. The two \Prs crystals used as quantum memory and spectral filter are bulk samples (Scientific Materials), with an active-ions concentration of $0.05\,\%$ and length $5\,\mathrm{mm}$ and $3\,\mathrm{mm}$, respectively. Then the retrieved photons pass through a band-pass filter (centered at $600\,\mathrm{nm}$, linewidth $10 \,\mathrm{nm}$, Semrock).
We further protect the SPD with a mechanical shutter which remains closed during the whole memory and filter preparation. The storage sequences are synchronized to the cycle of the cryostat ($1.4\,\mathrm{Hz}$) with a home-made synchronization circuit, with the aim of reducing the effects of mechanical vibrations.
To detect the retrieved single photons at $606\,\mathrm{nm}$ we use a Laser Components SPD with $50\,\%$ detection efficiency and $10\,\mathrm{Hz}$ dark-count rate. For auto-correlation measurements (section \ref{AC}) we add a second SPD (tau-SPAD, detection efficiency $45\,\%$, dark count rate $15\,\mathrm{Hz}$, PicoQuant).

\section{Spin-wave Atomic Frequency Comb Protocol}
\label{fullAFC}

\subsection{AFC preparation}

To reshape the absorption profile of the memory crystal (see Fig. 1 in the main text for the energy level scheme) into an atomic frequency comb we proceed as follows:
we start by preparing a wide transparency window by sending strong laser pulses (the maximum power being $21\,\mathrm{mW}$) and sweeping them by $16\,\mathrm{MHz}$. This has the effect of emptying the $1/2_g$ and $3/2_g$ states of a portion of atoms, creating a $18\,\mathrm{MHz}$-wide transparency window. We then send $4\,\mathrm{MHz}$ broad {\it burn-back} pulses resonant with the transition $5/2_g - 5/2_e$ to repump back atoms in the states $1/2_g$ and $3/2_g$. Afterwards we clean the spin-storage state, the $3/2_g$, using $5\,\mathrm{MHz}$ broad pulses resonant to the $3/2_g - 3/2_e$. We want this state to be as clean as possible, in order to reduce the noise generated by the control pulses (CPs) during the spin wave storage. At this stage, we have a $4\,\mathrm{MHz}$-wide single class absorption feature resonant to the transition $1/2_g - 3/2_e$, where we can prepare the atomic frequency comb (AFC).

\begin{figure}
   \centering
   \includegraphics[width=0.5\textwidth]{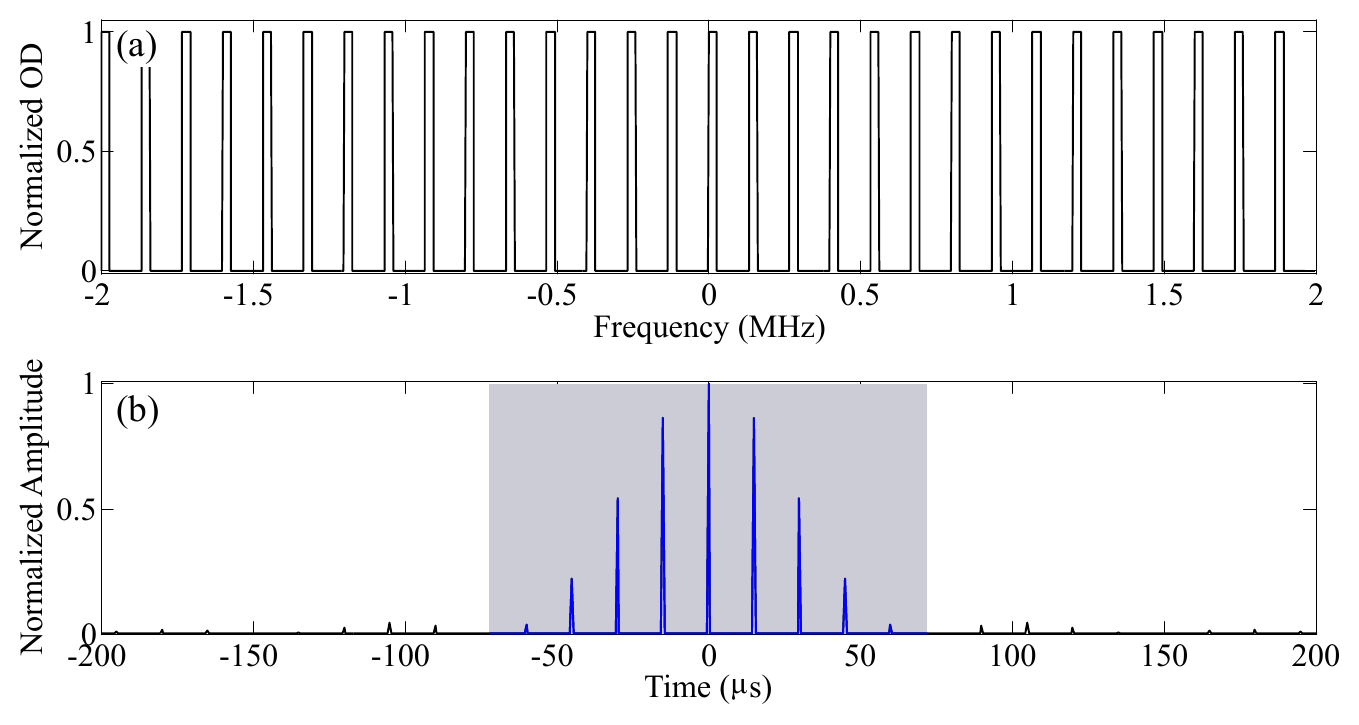}
   \caption{(a) Example of desired atomic frequency comb, with optimized finesse; (b) pulse train obtained by performing the Fourier transform of the comb shown in panel (a). The shaded rectangle shows the principal part of the waveform that we select.}
\label{fig:AFC}
\end{figure}

Finally we create the comb structure using a procedure inspired by \cite{Jobez2016}: we simulate the desired comb (an example being shown in Fig. \ref{fig:AFC}(a)), deciding the shape of the peaks (square), fixing the optical depth (OD) and the background optical depth ($d_0$), and calculating the best finesse for each particular AFC storage time ($\tau$). We perform the Fourier transform of the simulated comb and cut the principal part of it ($50-150\,\mathrm{\mu s}$, depending on the storage time), as evidenced in Fig. \ref{fig:AFC}(b) by the shaded rectangle, to maintain the duration of the preparation within the limit imposed by the cryostat cycle. This pulse is finally renormalized for the non-linear response of the double-pass AOM. We then optically pump the $1/2_g - 3/2_e$ transition with this pulse train 950 times (1100 times for $\tau > 9\,\mathrm{\mu s}$), and each time we clean the $3/2_g$ spin-state with a $10\,\mathrm{\mu s}$ long pulse  at the frequency of the $3/2_g - 3/2_e$ transition and chirped by $4.8\,\mathrm{MHz}$.

In order to reduce the noise generated by the CPs during the spin wave storage, after the preparation of the comb we send 100 CPs separated by $25\,\mathrm{\mu s}$ and, afterwards, other 50 with a separation of $100\,\mathrm{\mu s}$. If these {\it cleaning} CPs are too close to each other the atoms are coherently driven between the ground and the excited state and some might remain in the former. 

\begin{figure}[h]
\centering
\includegraphics[width=0.5\textwidth]{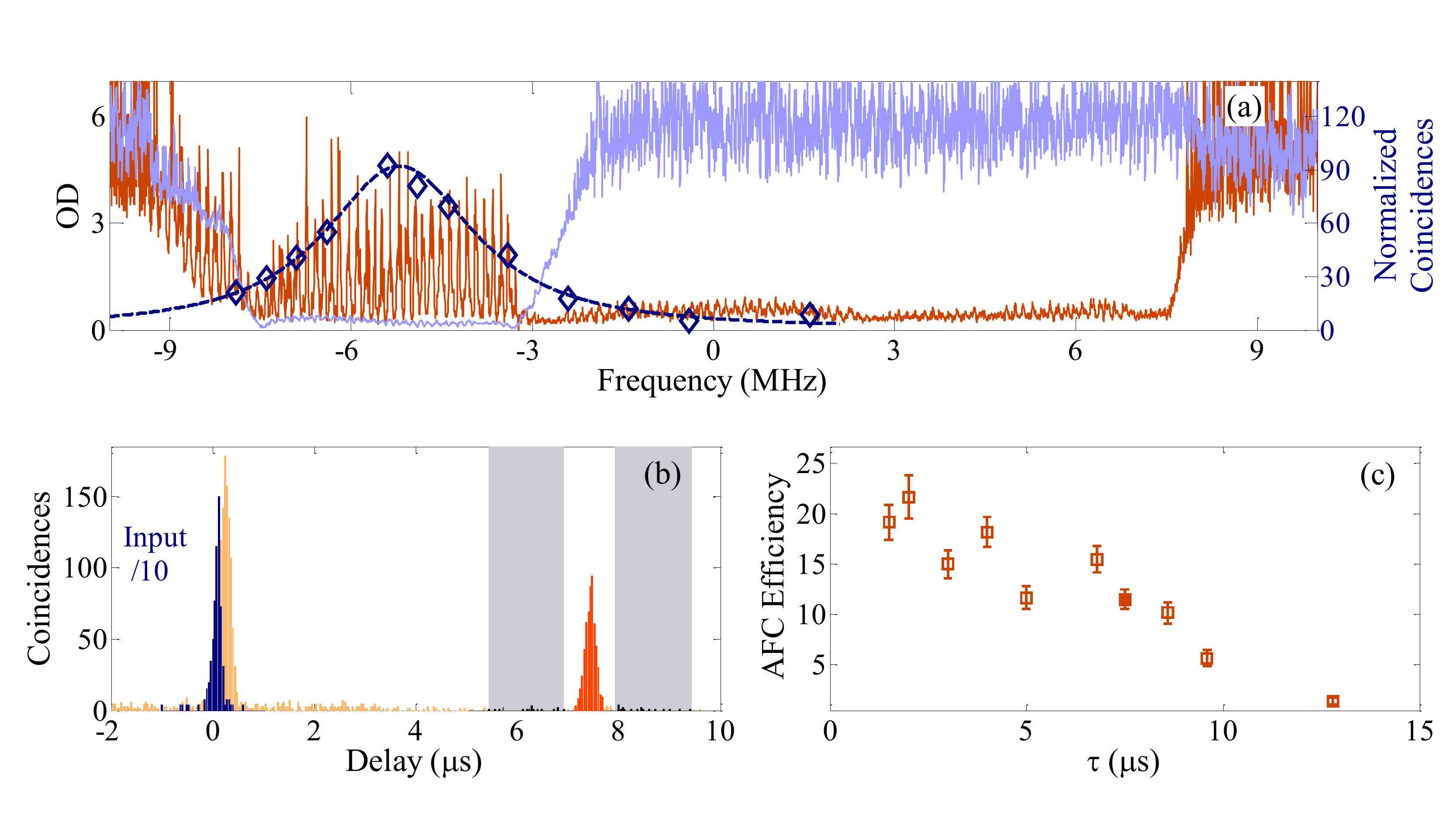}
\caption{(a) Example of atomic frequency comb prepared for $\tau = 7.3\,\mathrm{\mu s}$. The light violet trace is the transparency window that we burn in the filter crystal. The diamonds are signal-idler coincidence rates taken after preparing a single $800\,\mathrm{kHz}$ broad transparency window in the memory crystal and moving its frequency along the input photons. The error is smaller than the data points. The dotted blue line is a simulation of a Lorentzian peak with $FWHM = 2.8\,\mathrm{MHz}$ convoluted with a $800\,\mathrm{kHz}$-wide spectral hole. (b) Time histograms of the input photons (dark blue trace peaked at $0\,\mathrm{\mu s}$) and the AFC echo (red trace, including the transmitted signal). The shaded rectangles mark the areas used to measure the uncorrelated noise in the calculation of the $g^{2}_ {AFC,i}$ value. (c) AFC storage efficiency for different storage times, $\tau$. The filled square is $\tau = 7.3\,\mathrm{\mu s}$, whose coincidence histogram is shown in panel (b). }
\label{fig:2LE}
\end{figure}

In Fig. \ref{fig:2LE} we show a typical trace of atomic frequency comb prepared in the memory crystal and the trace of the spectral hole prepared in the filter crystal (panel (a)). The actual shape of the comb peaks is rather Gaussian than square, due to power broadening. Also shown is the spectrum of the heralded single photons reconstructed with signal-idler coincidence rates taken after preparing a single $800\,\mathrm{kHz}$ broad transparency window in the memory crystal and moving its frequency along the input photons. The dotted blue line is a simulation of a Lorentzian peak with $FWHM = 2.8\,\mathrm{MHz}$ convoluted with a $800\,\mathrm{kHz}$-wide spectral hole.

An example of coincidence histogram measured for $\tau = 7.3\,\mathrm{\mu s}$ and the AFC storage efficiency as a function of the storage time $\tau$  are shown in panels (b) and (c), respectively.

\subsection{Control Pulses}
\label{CP}
\begin{figure}[h]
   \centering
   \includegraphics[width=.45\textwidth]{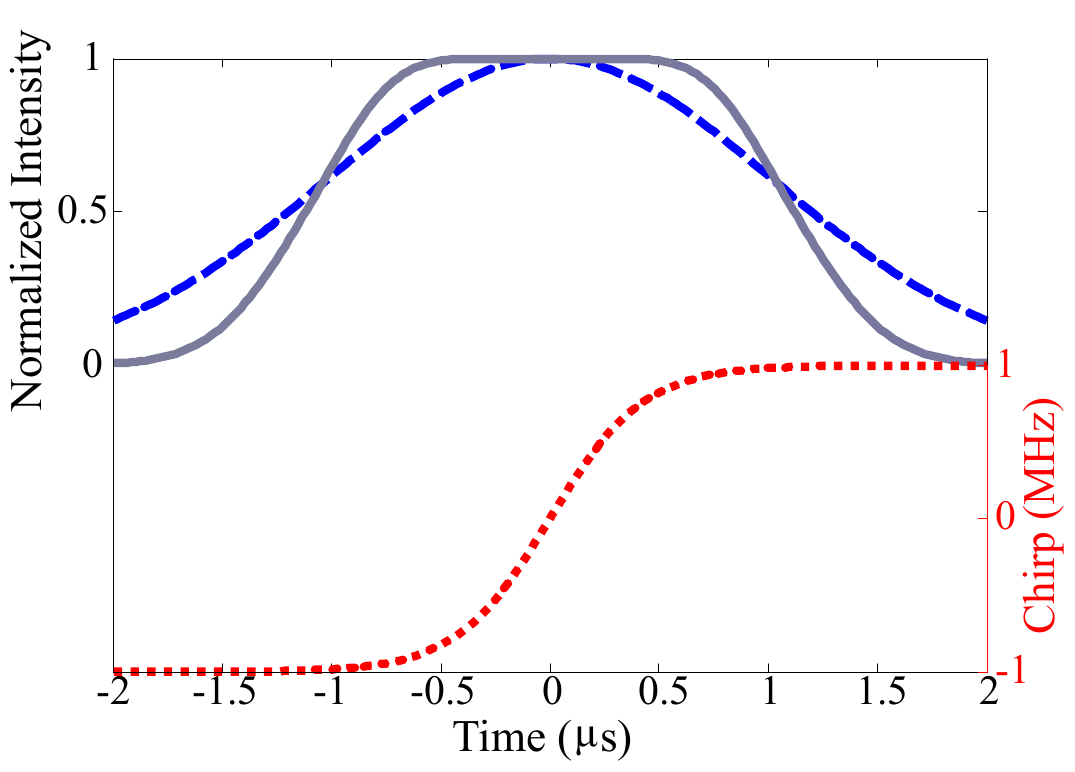}
   \caption{Control pulse waveform. The blue dashed trace is the waveform fed into the AOM. The resulting shape of the CPs looks like the gray solid trace. The CPs are chirped following the red dotted trace.}
\label{fig:CPs}
\end{figure}

The waveform that we use to generate our control pulses (CPs) is a Gaussian (blue dashed trace in Fig. \ref{fig:CPs}) with full-width at half maximum $FWHM = 2.4\,\mathrm{\mu s}$. As the AOM that we use to modulate the amplitude and frequency of the pulses has a non-linear response, the output waveform looks more squarish (gray solid trace). The FWHM, though, remains almost the same. This shape increases the efficiency of the CPs, so that we can take advantage of having shorter waveforms. This will be important in particular for the semi-conditional storage (section \ref{SCS}), in which we have to send the CPs as fast as possible after the heralding detection. In this case using longer waveforms means having the second CP closer to the echo, thus increasing the noise floor. 
These CPs are frequency chirped with a hyperbolic tangent (red dotted trace in Fig. \ref{fig:CPs}), that makes the transfer more effective around the central frequency \cite{Rippe2005}.

The transfer efficiency of the control pulses $\eta_T$ is measured by preparing an AFC and sending the first control pulse before the rephasing of the atoms: calling $\eta_{AFC}$ the efficiency of the AFC storage and $\eta'_{AFC}$ the efficiency of the AFC rephasing after the control pulse, we have:
$$\eta_T = 1 - \frac {\eta'_{AFC}} {\eta_{AFC}} = (72.5 \pm 1.3)\,\%.$$

\subsection{Noise generated by the control pulses}

\begin{figure}[h]
   \centering
   \includegraphics[width=.45\textwidth]{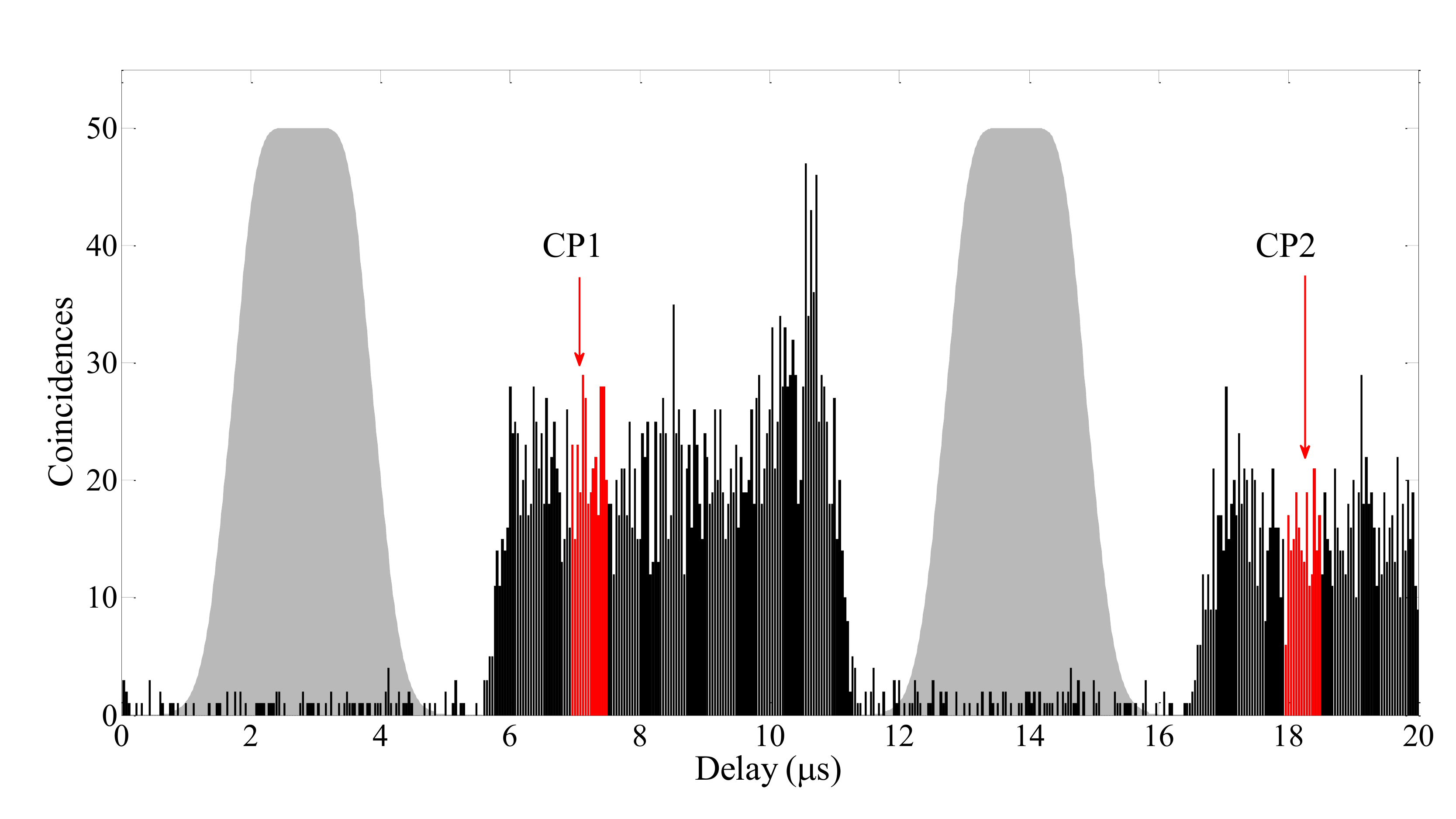}
   \caption{Trace of the noise generated by one control pulse (CP1 is the temporal mode of the AFC echo) and by both control pulses (CP2 being the position of the spin wave echo).}
\label{fig:NCP1CPs}
\end{figure}

We use a semi-conditional sequence (explained in more detail in section \ref{SCS}) to measure the noise generated by the first control pulse and by both of them. Each time that we detect an heralding photon we send a CP, then the temporal gate is opened in the position of the AFC echo and closed before the second CP is sent. We open it again and measure the noise generated by both pulses in the temporal mode of the spin wave echo (Fig. \ref{fig:NCP1CPs}). The noise generated by the first control pulse is higher than the noise after both CPs. Specifically the noise after the two CPs (red trace around $18\,\mathrm{\mu s}$) is $(2.3 \pm 0.1) \times 10^{-3}$ photons per storage trial, i.e. $86\%$ of the noise after the first CP (red trace about $7.3\,\mathrm{\mu s}$), measured in the position of the AFC.

If the $3/2_g$ spin state is not well cleaned, a portion of the remaining atoms depending on the transfer efficiency, is promoted to the excited state by the first control pulse, possibly giving rise to incoherent fluorescence. This excess of population at the excited state might then be coherently transferred to the ground state by the second control pulse, thus decreasing the contribution to the noise.

\subsection{Comb analysis}
\label{efficiencies}
The efficiency of the AFC echo can be predicted by analyzing the trace of the comb (Fig. \ref{fig:2LE}(c)) according to the model described in ref. \cite{Afzelius2009}. Assuming Gaussian peaks, we express the AFC internal efficiency with the following equation:
$$ \eta^{int}_{AFC} = \tilde{d} ^2 e^{-\frac{7}{F^2}}e^{-\tilde{d}}e^{-d_0},$$
where $\Delta = 1/\tau$ is the distance between the peaks of the comb, $\gamma$ is their full width at half maximum, $F=\Delta/\gamma$ is the comb finesse, $\tilde{d}=OD/F$ is the effective optical depth, and $d_0$ is the absorption background due to imperfect optical pumping.
The total efficiency that we measure is reduced by a factor $\eta_{BW} \approx 70\,\%$, due to the bandwidth mismatch between photons and comb (see section III of the main text).
Assuming for the comb reported in Fig. \ref{fig:2LE} a finesse $F = 3.8 \pm 0.3$, average optical depth and linewidth of the peaks $OD = 3.5 \pm 0.2$ and $\gamma = (36 \pm 3)\,\mathrm{kHz}$, respectively, and a background $d_0 = 0.27 \pm 0.1$, the expected total efficiency is $\eta_{AFC} = (11.3 \pm 1.4)\,\%$ which agrees very well with the experimentally measured $\eta_{AFC}^{exp} = (11.0 \pm 0.5)\,\%$. With the same comb parameters, the expected internal efficiency, assuming no loss due to the bandwidth mismatch, would be $\eta^{int}_{AFC} = (16\pm 2)\,\%$. 

The AFC efficiency can be separated into different contributions as follows:
$$\eta_{AFC} = \eta_{abs} \eta_{reph}\eta_{loss},$$
where $\eta_{abs}$ ($\eta_{reph}$) is the absorption (rephasing) efficiency of the comb. The factor $\eta_{loss}$ accounts for the loss due to absorption in the background. The absorption efficiency in the comb can be calculated as $\eta_{abs} = (1-e^{-\tilde{d}})\eta_{BW} = 43\,\%$. The rephasing efficiency of the comb is thus $\eta_{reph} = 34\,\%$. 

The spin wave efficiency can be analogously expressed as 
$$\eta_{sw} = \eta_{AFC}\eta_{T}^2\eta_C$$
assuming the transfer efficiency $\eta_{T}$ be the same for both control pulses and the spin state exhibit a Gaussian inhomogeneous broadening leading to a decoherence effect (showed in Fig. 3(a) of the main manuscript) quantified by $\eta_{C} = exp\left(-\frac{(T_s\cdot\Gamma_{inhom})^2}{2\cdot log(2))}\cdot \pi^2\right) = 87.3\,\%$ \cite{Afzelius2010,Gundogan2013,Gundogan2015}. 
The expected value of the spin wave efficiency is $\eta_{sw} = (5.2\pm0.6)\,\%$. The small mismatch with the experimentally measured one, $\eta_{sw}^{exp} = 3.6\pm0.2\,\%$, can be due to the fact that our chosen coincidence window, $\Delta T_d$, contains about $80\,\%$ of the echo and that the filter crystal might have residual background due to imperfect optical pumping. 

Given the control pulse efficiency $\eta_{T} = 72.5\,\%$ (see section \ref{CP}), we define a spin wave write and read-out efficiency as $\eta_{W} = \eta_{abs}\eta_{T}$ and $\eta_{RO} = \eta_{T}\eta_{reph}$, respectively (thus rewriting the total spin-wave efficiency as $\eta_{sw} = \eta_{loss}\eta_{W}\eta_C\eta_{RO}$), which we estimate to be $31\,\%$ and $24\,\%$, respectively. 
Taking this into account and considering that the first control pulse gives a measured noise floor of $(2.3 \pm 0.1) \times 10^{-3}$, we infer non-classical correlation between the single telecom photons and the collective spin excitations during the storage with a $g^{(2)}_{sw,i}$ of the order of $20$.

\section{Storage Sequences}
\label{sequences}
After the preparation of the comb we run different sequences for the storage of single photons. In one of them we try to store without knowing in advance if there is a single photon at the quantum memory, we call this sequence {\it unconditional}. In the second sequence, that we call {\it semi-conditional}, we condition the first storage trial on the detection of the heralding photon (idler), followed by further unconditioned trials.

\subsection{Unconditional sequence}
\begin{figure}[h]
   \centering
   \includegraphics[width=0.5\textwidth]{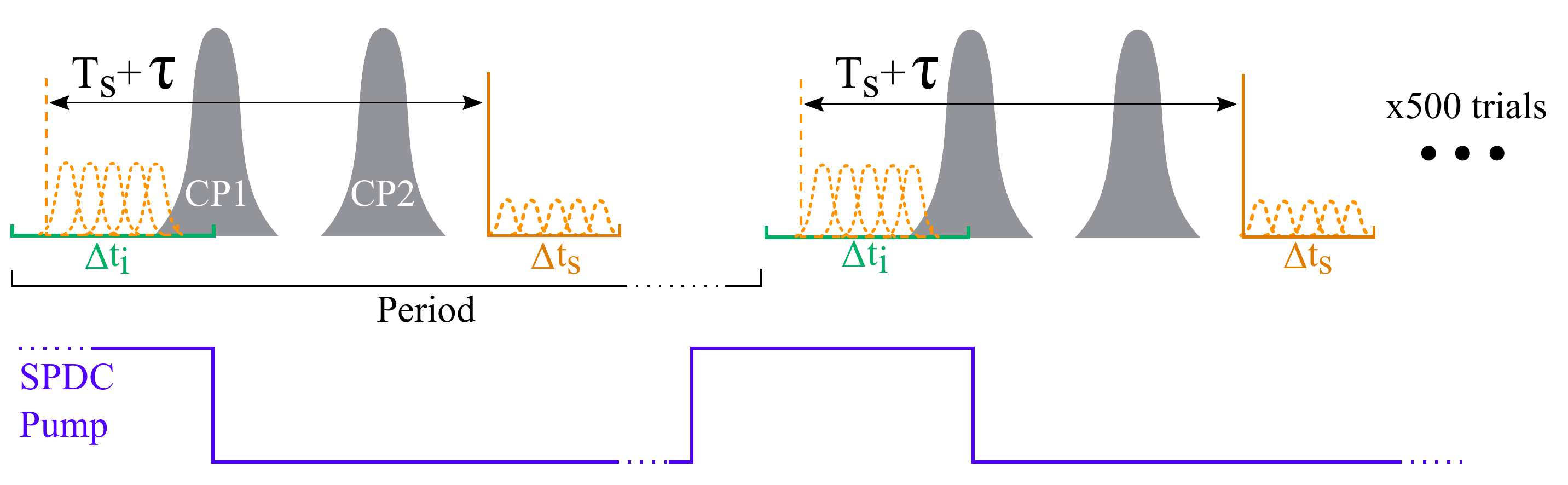}
   \caption{Unconditional sequence: storage of single photons not conditioned on the detection of the heralding photon. The purple line shows the voltage sent to the AOM in front of the SPDC cavity, thus the switching on and off of the pump laser.}
\label{fig:non-c_seq}
\end{figure}

In this sequence, sketched in Fig. \ref{fig:non-c_seq}, after the preparation of the memory, we perform 500 storage trials, separated by $190\,\mathrm{\mu s}$. We choose this time, longer than the relaxation time of the excited state, in order to reduce accumulated noise in the spin wave echo mode due to the multiple storage trials. Each storage trial consists of two transfer pulses (write and read), separated by $T_s = 6\,\mathrm{\mu s}$. 
The gate of the idler detector is opened for a time $\Delta t_i = 6\,\mathrm{\mu s}$ and it closes when the write pulse arrives to its maximum intensity. For this measurement, we assemble a Hanbury-Brown Twiss setup (fiber BS in Fig. \ref{fig:setup}) after the memory such that we can reconstruct both the cross-correlation $g^{2}_{swe,i}$ between the retrieved photons and the heralding photons and the auto-correlation of the retrieved photons. The two detectors of the signal are switched on $1.3\,\mathrm{\mu s}$ after the read pulse and are maintained open for a time $\Delta t_s = 4.5\,\mathrm{\mu s}$. 
Together with the closure of the idler gate, an RF-signal is sent to the AOM of the pump laser of the source in order to keep it off for $30\,\mathrm{\mu s}$ during the detection of the signal photons (purple line in Fig. \ref{fig:non-c_seq}).

\subsubsection{Cross-correlation}
\label{CC}

\begin{figure}[h]
   \centering
   \includegraphics[width=0.5\textwidth]{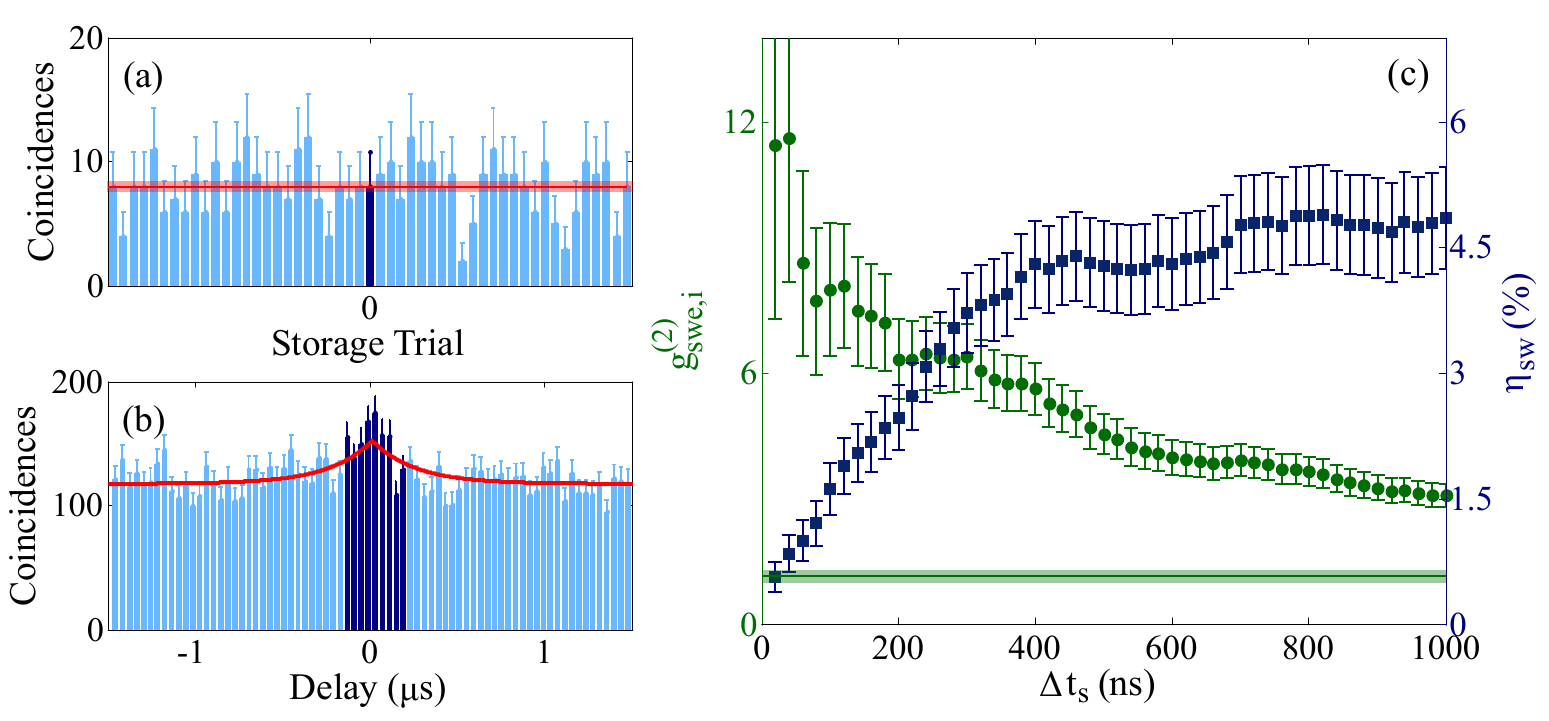}
   \caption{Panel (a): Second order autocorrelation histogram of the stored and retrieved single photons $g^{(2)}_{swe,swe}$. The darker bar indicates the coincidence counts in a detection window $\Delta t_d = 320 \,\mathrm{ns}$ about $0$ delay in the same storage trial. Panel (b): Second order autocorrelation histogram of the idler photons measured in CW configuration. The region marked with darker bars indicate the integration window $\Delta t_d = 320 \,\mathrm{ns}$. Panel (c): Second order cross-correlation between idler and retrieved signal (circles) and spin-wave echo efficiency (squares) as a function of the integration window, $\Delta t_d$.}
\label{detwindow}
\end{figure}

The cross-correlation between the retrieved signal and idler photons is defined as $g^{(2)}_{swe,i}= \frac{P_{swe,i}}{P_{swe} P_i}$, where $P_{swe,i}$ is the coincidence probability between the idler and the retrieved signal photons, while $P_{swe}$ ($P_i$) is the retrieved signal (idler) uncorrelated probability. We reconstruct the $g^{(2)}_{swe,i}$ value as the ratio between the coincidences detected at a storage time $\tau+T_S = (7.3+6)\,\mathrm{\mu s} = 13.3\,\mathrm{\mu s}$ within the same storage trial and the average of the coincidences between signal and idler photons detected in the spin wave temporal mode of the 20 neighboring storage trials (a typical result being shown in Fig. 2(b) of the manuscript). In both cases the integration window is $320\,\mathrm{ns}$.
However, we also analyze the $g^{(2)}_{swe,i}$ value, along with the spin wave efficiency as a function of the integration window. The results are shown in Fig. \ref{detwindow}(c). 

\subsubsection{Auto-correlation}
\label{AC}
The auto-correlation of the retrieved signal is defined as $g^{(2)}_{swe,swe}=\frac{P_{swe1,swe2}}{P_{swe1} P_{swe2}}$, where $P_{swe1,swe2}$ is the probability to have coincidences between the two output ports of the signal arm, while $P_{swe1}$ ($P_{swe2}$) is the uncorrelated probability to detect a photon in the output 1 (2) of the Hanbury-Brown Twiss setup. We calculate the unconditional autocorrelation as the ratio of the coincidences between the signals detectors in the same storage trial and the coincidences in the previous and following 10 storage trials. As the signal can arrive everywhere in $\Delta t_s$ thanks to the multimodality of the protocol, we build the coincidence histogram between the signal photons at the two outputs of the fiber BS considering the whole detection window. Then, to calculate the auto-correlation $g^{(2)}_{swe,swe}$, we integrate in a window of width $\Delta T_d = 320\,\mathrm{ns}$ around $0\,\mathrm{\mu s}$ delay, to be consistent with the cross-correlation measurement. The result is shown in Fig. \ref{detwindow}(a) and the value that we extract is $g^{(2)}_{swe,swe}= 1.0 \pm 0.4$. The large error bar is due to the low statistics. 
This value is remarkably lower than expected for a thermal state produced by a SPDC process \cite{Tapster1998}. 
However, we measure the second order autocorrelation of the signal photons before the memory to be $g^{(2)\, mm}_{swe,swe} (0)= 1.18 \pm 0.02$, for $3.9$ effective spectral modes \cite{Rielander2016}. From this value, we can extrapolate the autocorrelation of the input photons for the single spectral mode case, which would measure $g^{(2) \, sm}_{swe,swe} (0)= 1.70 \pm 0.03$.
After the memory, the signal autocorrelation will be affected by the noise produced by the storage protocol. We quantify this contribution with the unconditional signal-to-noise ratio, i.e. the probability to detect a signal photon after the retrieval over the noise in unconditional measurements. We plug this value, $SNR^{unc} = (P_{swe}-P_n)/P_n = 0.030 \pm 0.003$, in the theoretical model described in \cite{Rielander2014}, which assumes that the noise is not bunched, and find that after the noisy storage protocol the expected signal autocorrelation is $g^{(2) \,sws}_{swe,swe} (0)= 1.01 \pm 0.08$. Finally, the finite integration window also contributes to decrease the autocorrelation \cite{Rielander2016} to $g^{(2) \,th}_{swe,swe} (320 \,\mathrm{ns})= 1.0\pm 0.1$, which agrees with the experimental value ($g^{(2)}_{swe,swe} (320 \,\mathrm{ns})= 1.0 \pm 0.4$). This also suggests that the noise is indeed not bunched.

For the idler photons we measure the unconditional autocorrelation in a CW configuration because no storage is involved \cite{Rielander2016}. A typical measurement for a pump power of $4.3\,\mathrm{mW}$ is shown in Fig. \ref{detwindow}(b) from which we calculate a value of $g^{(2)}_{i,i} (320\,\mathrm{ns})= 1.32 \pm 0.04$.  

\subsubsection{Cauchy-Schwarz inequality}
\label{CS}

To confirm the quantum correlation between the idler and the stored and retrieved signal photons, we calculate the Cauchy-Schwarz parameter $R = \frac{({g^{(2)}_{swe,i}})^2}{g^{(2)}_{swe,swe} \cdot g^{(2)}_{i,i}}$, which is expected to be lower than 1 for classical fields. Using the values reported in sections \ref{CC} and \ref{AC} we calculate $R = 28\pm 12$, which exceeds the classical benchmark by two standard deviations. The big errorbar is due to the low statistics in the second order autocorrelation measurements of the retrieved signal photons. Considering a wider integration window, e.g. $\Delta T_d = 1\,\mathrm{\mu s}$, the uncertainty decreases due to better statistics. The auto- and cross-correlation values also decrease due to the contribution of uncorrelated noise. Nonetheless, the Cauchy-Schwarz inequality is still violated, as summarized in Table \ref{CStable}.

\begin{table}
\begin{center}
\caption{Second order auto- and cross-correlation values of idler and retrieved signal photons as obtained from the experimental histogram when considering different integration windows. The corresponding Cauchy-Schwarz parameters R are also reported.} 
\label{CStable}

\begin{tabular}{|c|c|c|}
\hline $\Delta T_d $ & $320\,\mathrm{ns}$ &  $1\,\mathrm{\mu s}$ \\

\hline

$g_{swe,swe}^{(2)}$ & $1.0 \pm0.4$ & $1.1 \pm 0.2$ \\

$g_{i,i}^{(2)}$&$1.32 \pm 0.04$&$1.14\pm 0.02$ \\

$g_{swe,i}^{(2)}$&$6.1 \pm 0.7$&$3.1 \pm 0.3$ \\

$R$&$28 \pm 12$&$8.3 \pm 2.3$ \\

 \hline
 \end{tabular}
\end{center}

\end{table}

Note that, as demonstrated in the calculation of the expectation value of section \ref{AC}, the biggest contribution to the signal autocorrelation is given by the noise and that this is not bunched. Thus, any modification in the measurement that increases the noise (e.g. the use of the semi-conditional sequence described in section \ref{SCS}), would only further approach the autocorrelation to the value of 1, thus lowering the classical threshold.

\subsection{Semi-conditional sequence}
\label{SCS}
The drawback of the unconditional sequence is the low count rate, resulting in extremely long integration times. This is due to the fact that the measurement duty cycle is low in this configuration ($\sim 2\,\%$ with respect to the measurement period). 
We thus set a semi-conditional sequence which allows us to estimate the signal-idler cross-correlation, $g^{(2)}_{swe,i}$, in considerably shorter measurement times.

\begin{figure}[h]
   \centering
   \includegraphics[width=0.5\textwidth]{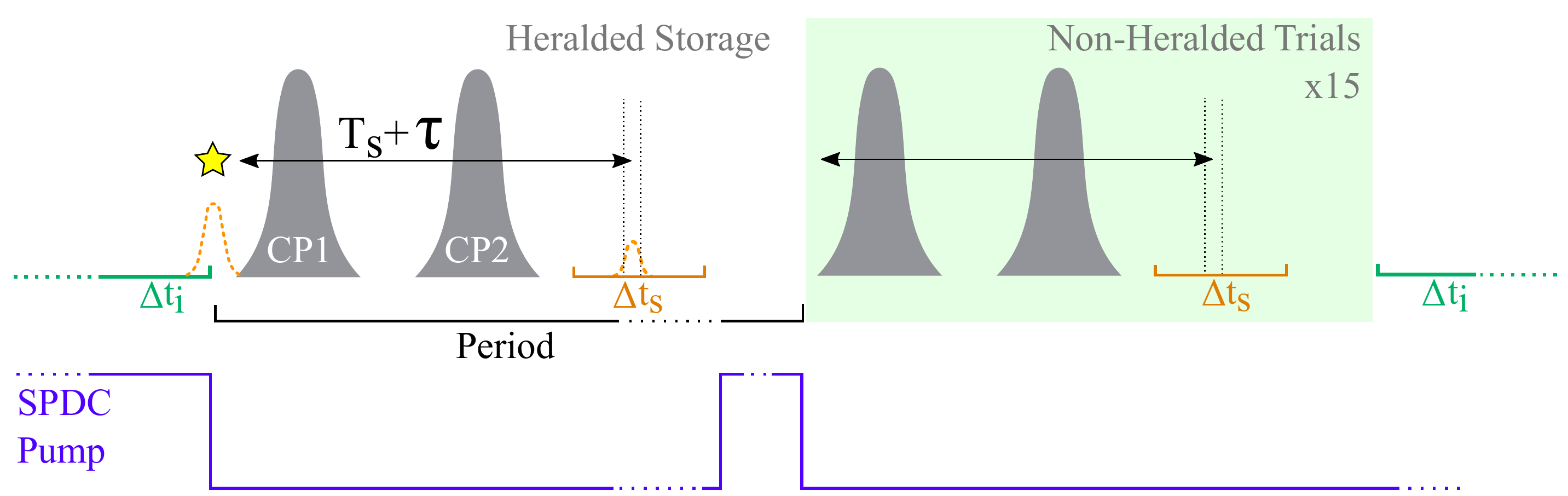}
   \caption{Semi-conditional sequence: storage of single photons conditioned on the detection of the heralding photon. After each conditional storage we perform other 15 unconditional storage trials. The purple line shows the voltage sent to the AOM in front of the pump, thus the switching on and off of the pump laser.}
\label{fig:s-c_seq}
\end{figure}

\begin{figure}[h]
   \centering
   \includegraphics[width=0.45\textwidth]{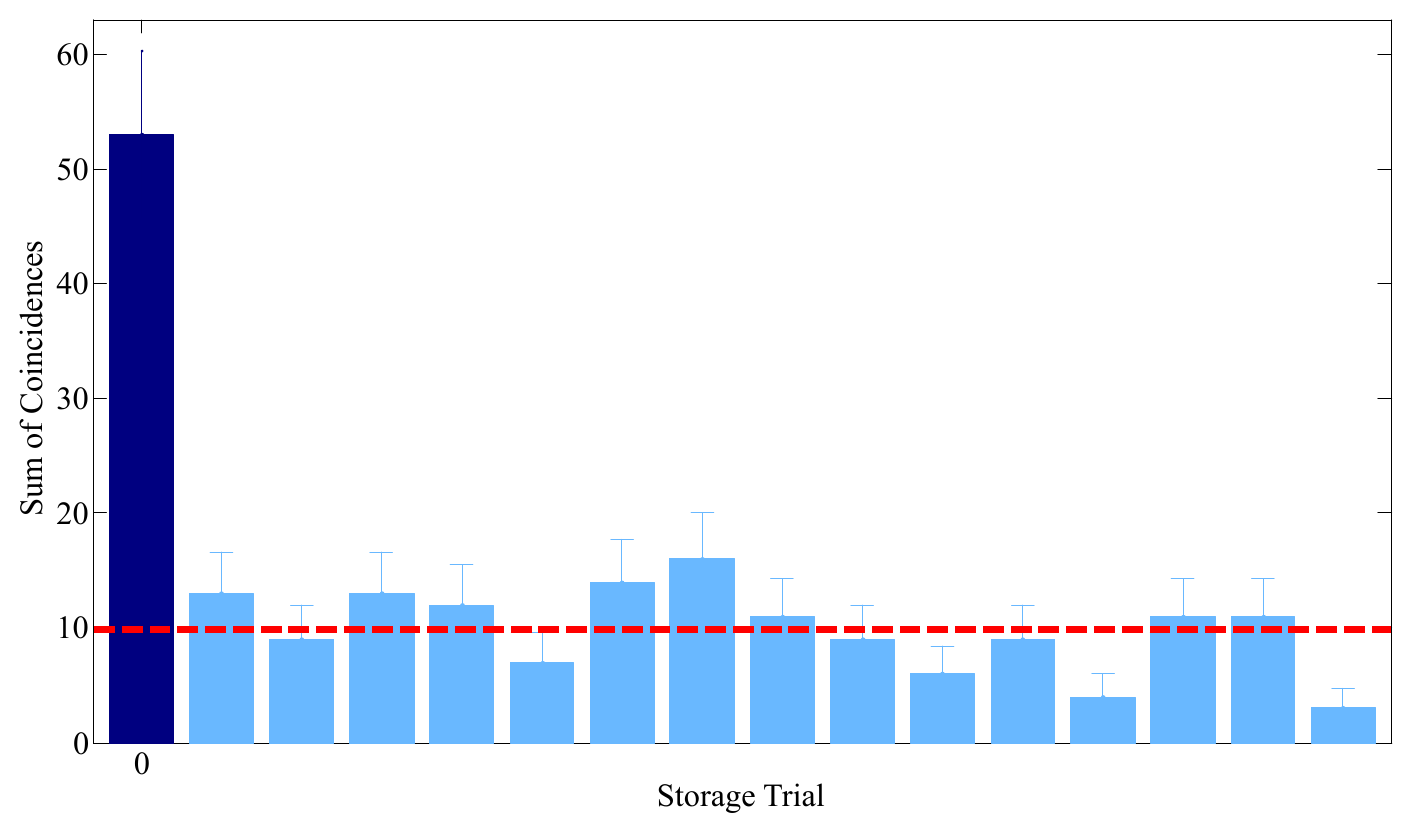}
   \caption{Cross-Correlation measurement for a spin wave storage $T_s = 6\,\mathrm{\mu s}$, using the semi-conditional sequence: the bar at 0, in dark blue, represents the sum of the coincidences in a $320\,\mathrm{ns}$ window placed $\tau + T_s = 13.3\,\mathrm{\mu s}$ after the heralding photon. The other bins represent coincidences (for the same coincidence window size and position) in the following 15 storage trials, the red dashed line being the average value. }
\label{fig:cc}
\end{figure}

In this sequence, sketched in Fig. \ref{fig:s-c_seq}, after the preparation of the memory, we open the idler gate and continuously (every $80\,\mathrm{ns}$) check for heralding events. Each time that we detect an idler photon we close the gate of the telecom detector, we switch off the SPDC pump (for $40\,\mathrm{\mu s}$), and we start the storage cycle. This consists of two identical control pulses, the write and the read, sent with a relative delay of $T_s$. After the second pulse, the signal gate is opened for a time $\Delta t_s$ (see Fig. \ref{fig:s-c_seq}) to detect the retrieved photon. We know that the retrieved photon will arrive at a time $\tau+T_s$ after the heralding. To estimate the noise, after the retrieval we send 15 pairs of CPs (unconditional) with a period of $190\,\mathrm{\mu s}$. In this case the photon is coupled in a single mode fiber and measured with one SPD.
Analogously to the unconditional case, we calculate the cross-correlation between the retrieved signal and idler photons as $g^{(2)}_{swe,i}= \frac{P_{swe,i}}{P_{swe} P_i}$. Again we consider the ratio of the coincidences between the heralds and the retrieved photons at $\tau+T_S$ in the same storage trial, and the average of the coincidences in the following 15 storage trials. The resulting coincidence histogram is reported in Fig. \ref{fig:cc}, to be compared with the unconditional coincidence histogram shown in Fig. 3(a) of the main manuscript. The $g^{(2)}_{swe,i}$ for the semi-conditional measurement is $5.0\pm0.3$.
Note that in the same conditions (pump power and storage time), the semi-conditional sequence provides a slightly lower $g^{(2)}_{swe,i}$ value, mainly due to the higher noise floor, $(2.0\pm0.1)\times 10^{-3}$, with respect to the unconditional sequence.

This sequence has been employed to measure $g^{(2)}_{swe,i}$ for different $T_s$ (whose results are shown in Fig. 4 of the main text).

\begin{figure}[h]
   \centering
   \includegraphics[width=0.45\textwidth]{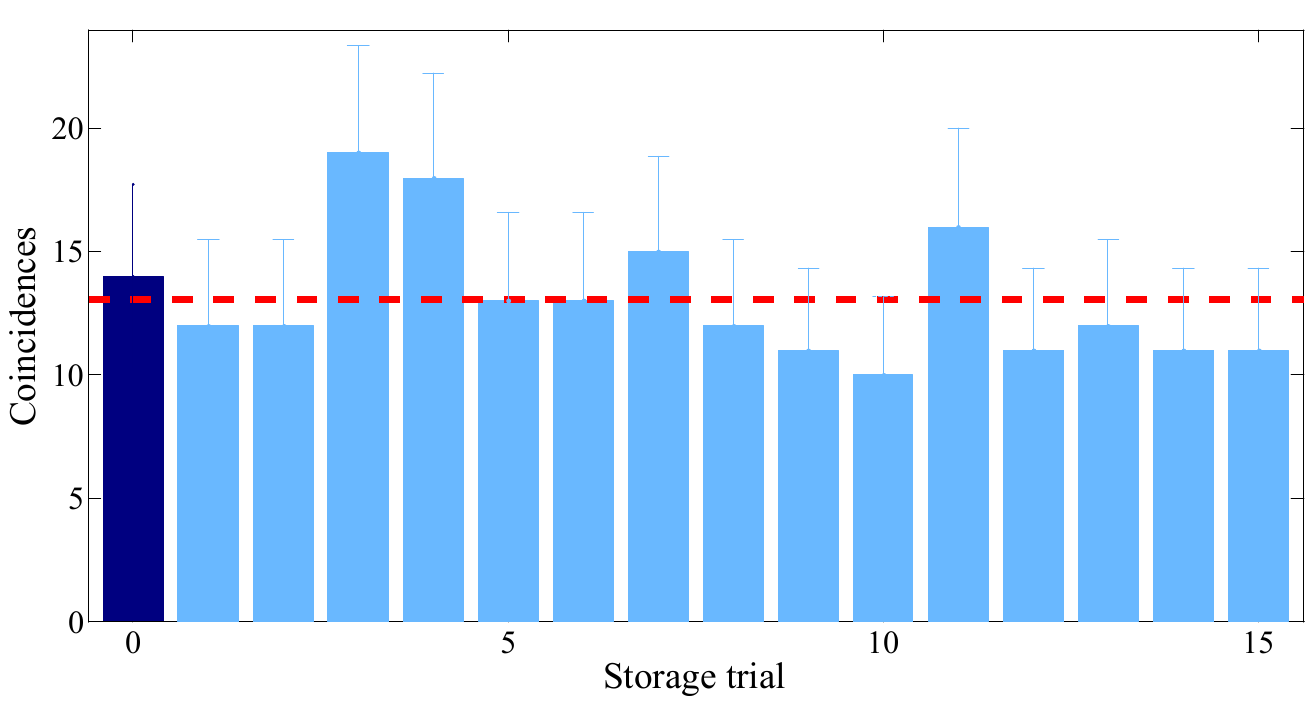}
   \caption{Cross-Correlation measurement of the noise for a spin wave storage $T_s = 6\,\mathrm{\mu s}$, using the semi-conditional sequence. The measurement of Fig. \ref{fig:cc} is repeated blocking the signal photons in front of the quantum memory.
}
\label{noise}
\end{figure}

We also repeated the same measurement while blocking the signal photons before the quantum memory with a beam block. In this way we can measure the cross-correlation of the noise, which is $g^{(2)}_{n,i}(320\,\mathrm{ns})= 1.1\pm0.3$ ($g^{(2)}_{n,i}(3\,\mathrm{\mu s})= 1.0 \pm 0.1$), as shown in Fig. \ref{noise}.

\end{document}